\documentclass[preprint,12pt]{elsarticle}

\usepackage{amssymb}

\journal{Advances in Quantum Chemistry}

\begin{document}
\title{Short time-to-solution Quantum Monte Carlo for catalysed hydrogen synthesis. Tools give CO hydrolysis activation barriers to 1kJ/mol on Pt(111).}

\author{Ali Ba{\u g}c{\i}}
\affiliation{Computational and Gravitational Physics Laboratory, Department of Physics, Faculty of Science, Pamukkale University, Denizli, Turkey.}
\author{Philip E. Hoggan${}^{*}$}
\affiliation{Institut Pascal, UMR 6602 CNRS, BP 80026, 63178 Aubiere Cedex, France.}

\begin{abstract}
Hydrogen synthesis is a clean, sustainable alternative to fossil fuel \cite{gals}. It has come of age:
prototyping various aspects of hydrogen power are hot topics.

In 9 out of 10 reactions, a solid catalyst is used. Here hydrogen production (via water-gas shift) is studied. Adsorbed reactants are optimidsed on model Pt(111). Focus is on partial O-H bond dissociation, when CO is co-adsorbed with water on this plane. hydrogen is the product.
Many chemical reactions involve bond-dissociation. This process is often the key to rate-limiting reaction steps at solid surfaces.

Bond-breaking is poorly described by Hartree-Fock and DFT methods, our embedded active site approach is used.
We showcase Quantum Monte Carlo (QMC) methodology using the ground-state Slater Determinant of a simple four primitive-cell layer model, oriented to expose Pt (111), to initialise the QMC.
This stochastic approach solves the Schr{\"o}dinger equation. It recently came of age for heterogeneous systems involving solids.
During hydrolysis of carbon monoxide, initial O-H bond stretch is rate-limiting. Its dissociation energy is offset by surface Pt-H bond formation.
The reactive formate (H-O-C=O) species formed by initial hydrolysis of CO, also interacting with a vicinal Pt.
The products are hydrogen (CO$_2$ by-product is mineralised. A H-atom dissociates from the formate, another is desorbed from Pt(111).  This yields pure hydrogen.

Single-determinant work with a novel averaging procedure is compared to a high-level configuration interaction (CI) wave-function. Activation barriers are given to 0.86kJ/mol (c.f. 0.7 of the CI benchmark). Active sites embedded in metal lattice (111) faces. These trial wave-functions guide QMC.
\end{abstract}
\maketitle
\section{Introduction} \label{introduction}

This work draws on exploratory results already published by us in \cite{gals,haqc92}, and the previous version on arXiv. The accuracy relies on treating a transition-state and asymptote structure comprising the same atoms. This transition state geometry is obtained from Variational Monte Czrlo force-constants.
In fact, asymptotic adsorbed species can be well-described by a ground-state single determinant wave-function. In this case the asymptote shows chemisorbed CO, in a 'tripod' geometry at a hollow site on Pt(111) binding to by 3 Pt-atoms that form an equilateral triangle in the close-packed 2D hexagonal mesh,with an approaching physisorbed water molecule. We thus have a simple wave-function, used to guide reference Quantum Monte Carlo (QMC) configurations, obtaining  0.5 kJ/mol standard error. It compares well to a fully pre-correlated benchmark \cite{hjcp}.

The subtlety of this work lies in how multi-configurational transition state QMC can be mimicked by two randomized runs continuing from the initial 5000 data-points which lead to 2kJ/mol standard error in activation barriers. Two runs were extended separately, each over a further 5000 data-points. These, in turn were distributed over 2048 cores each, with a target-weight of 16384 (or, selected fluctuations with branching and dying as well as drift around 8 configurations per core). This alone is enough to reduce standard error to below 1.6 kJ/mol per run, on this notoriously difficult structure to represent with a short configuration expansion.

Hydrogen powered trains, buses and cars are now in use and described in the popular press \cite{SA}.
Jet engines tested on hydrogen by Rolls Royce funded by airlines show the only remaining difficulty is storage requiring 4 times the space of kerosene, and condensation \cite{bbc}. In the current context of attempting to limit climate change, fossil fuel combustion is scrutinized due to greenhouse gas concerns: hydrogen production has come to the forefront.

Hydrogen can be obtained by electrolysis of water, which requires two O-H bond breakages. The water-gas shift (wgs) reaction is as follows \cite{absi}:
\begin{equation}
CO + H_2 O \rightarrow  CO_2 + H_2
\end{equation}
Carbon monoxide (toxic effluent), when hydrated gives hydrogen (N.B. the CO$_2$ can be mineralised as carbonate and removed).
CO, pre-adsorbed on metal surfaces, is approached by a physisorbed water molecule. Few metals with modest interaction strength with CO are suitable (e.g. copper interacts too strongly, becoming inert (poisoned),  but nickel and platinum are used). Close-packed (111) faces are the most active, when available, because most of these metals have an fcc unit mesh. Adsorbed CO is polar (C$^{\delta +} $) and reactive towards hydration (unlike free CO). The reaction with water begins by stretching one O-H bond, then the catalyst surface stabilises the product, binding it at a different metal atom. A simple overview is: O=C-(OH)-Pt-Pt-H, where a vicinal Pt atom forms a Pt-H bond. It effectively has much lower activation energy (than water O-H dissociation). Initial water attack of CO at Pt(111) is the rate-limiting step, studied in what follows. Of course, it is not perfectly 'green' since the catalysts are often based on heavy metals that must be mined and purified. Nevertheless, a catalyst can function without poisoning for an enormous number of cycles (infinite re-use limit). Some can be cleaned and the hydrogen synthesis is a 'grey' but non-polluting scheme.

\section{Input structures (unit cells) molecules and TS@Pt(111)}

This section requires several comments on the model catalyst. Defining a super-cell that is 4 mesh thick compares well to the benchmark (using a 5 mesh thick slab for identical top and bottom), because of periodic continuation in the solid. This procedure is simply more convenient with multiples of 4, which is thick enough to avoid surface (skin) 2D-state effects.
\newpage
{\bf Asymptote, adsorbed CO, physisorped water}
\vskip4mm
\begin{figure}
\includegraphics[scale=0.8]{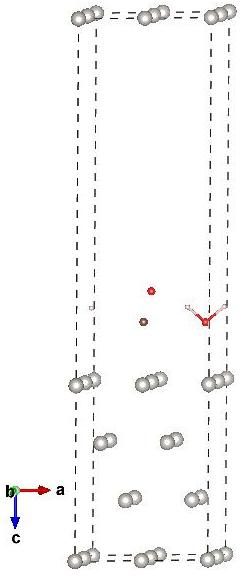}
\caption{Primitive unit cell for Pt, exposing (111) surfaces top and bottom. The lattice parameter a, is 3.92\AA \hskip2mm CO is adsorbed in Pt$_3$ hollows.}
\end{figure}
\vskip4mm
In this work, we compare the TS, from single determinant input to the multi-reference benchmark.

{\bf Transition state, from Variational Monte Carlo force-constants}
\begin{figure}
\includegraphics[scale=1.0]{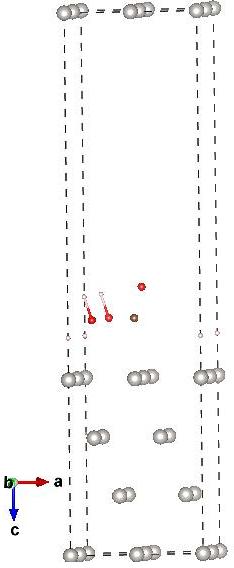}
\caption{Primitive unit cell for Ni, exposing (111) surfaces top and bottom. The lattice parameter a, is 3.92\AA \hskip2mm TS closeup (note H-atom at x=y=0.)}
\end{figure}

\vskip2cm
\section{Generic Jastrow factor}

A detailed account is given in  \cite{haqc92}, and references therein.

The software used, with the Generic Jastrow factor as default option is CASINO \cite{cas2020}.

In the present work, the Jastrow factor is fully optimised, and we have obtained convergence for variance followed by energy Variation Monte Carlo (VMC).
This Jastrow factor comprises three terms. The first term is made up of polynomials, with variational coefficients, as a function of each electron-electron distance for populations of instantaneous configurations. The second is the corresponding polynomial for electron-nuclear distance (and is distinct for each atom of the super-cell). The third term, which is indispensable, is for electron pair and nucleus (three body) nested sums over e-e and e-n distance, for each atoms.
High-order polynomial expansions in the instantaneous inter-particle distance (order 9 for 2 particle and order 4 for 3 particle) lead to large parameter sets (550 here), since individual atoms are treated.

\section{Results} \label{results}

The super-cells of 25 atoms (including 5 layers of 4 Pt atoms each, forming a lozenge primitive hexagonal mesh). This is completed by various CO and water geometries, which must be fully correlated by Variational Monte Carlo (VMC) optimisation of a generic Jastrow factor described in section 3, above and in  \cite{haqc92}. The resulting total energy for the asymptote is -462.0632 kJ/Mol

Combining all these elements, in a Fourier-transformed twist average gives the highly accurate result: standard error on the two TS-twists: 0.704 kJ/mol.
The total, is then 0.86 kJ/mol standard error on the barrier heights. Furthermore, the total energies of the two runs are -462.03664 and -462.03668 Ha giving a difference of less than 0.4 kJ/mol and average of -462.0365 which provides an estimate of the activation barrier as 70.1 kJ/mol, consistent with the benchmark value of 71 $\pm 0.7$ kJ/mol \cite{hjcp}.
\vskip4mm
\begin{tabular}{|c|c|c|c|c|c|c|}
\hline
steps   & E$_{ts}$ & max. & corr/rl & se mHa & err. in se & E$_a$  kJ/mol \\
\hline
a $5600$ & $-462.03672$ & -462.0362 & 84  64  & 0.93  & 0.0735 & $70.2$ $\pm 2.2$ \\
\hline
r1 $10449$ &  $-462.03664$ & -462.0365 & 46 32 & 0.62 & 0.04 & pop $16392$ \\
\hline
r2 $10975$ &  $-462.03668$ & -462.0365 & 46 32 & 0.59 & 0.04 & pop $16375$ \\
\hline
\hline
overall & asy & -462.0632 & se 0.86 & kJ/mol & barrier & 70.11 $\pm 0.86$ \\
\hline
\end{tabular}
\vskip4mm
This is very close to the numerical results of our previous single-determinant work, but with low standard error, bringing it well within the limits of chemical accuracy, at lower than 1kJ/mol error bars. It is also close to the benchmark for this hydration, rate-limiting barrier of CO hydrolysis on Pt(111), which remains our reference. Note that one standard deviation amounts to 1.1 kJ:mol for these two barrier estimates, therefore the present value of 70.1 is not significantly lower than the reference of 71 kJ/mol.

\vskip4mm
{\bf Acknowledgements.}
We thank GENCI for a special allocation on Irene (nov 2022 to nov 2023 CEA, Bruy\`eres-le-Ch\^atel, F) and renewals until 2025, as well as University of Clermont Mesocentre (data preparation, tests).

\end{document}